# Dielectric and multiferroic behavior in a Haldane spin-chain compound $Sm_2BaNiO_5$ due to an interplay between crystal-field effect and exchange interaction

**Sanjay Kumar Upadhyay, Kartik K Iyer and E.V. Sampathkumaran***

*Tata Institute of Fundamental Research, Homi Bhabha Road, Colaba, Mumbai 400005, India*

Abstract

The Haldane spin-chain (S=1) insulating compound, $Sm_2BaNiO_5$, which has been proposed to order antiferromagnetically around ($T_N =$) 55 K, was investigated for its complex dielectric permittivity, magnetodielectric and pyrocurrent behavior as a function of temperature ($T$). In order to enable meaningful discussions, the results of ac and dc magnetization and heat-capacity studies are also reported. We emphasize on the following findings: (i) There is a pyrocurrent peak near $T_N$, but it is shown not to arise from ferroelectricity, but possibly due to 'thermally stimulated depolarization current', unlike in many other members of this rare-earth series in which case ferroelectric features were reported at or above $T_N$; (ii) however, the pyrocurrent measured in the presence of a bias electric field (after cooling in zero electric field) as well as dielectric constant reveal a weak peak with increasing $T$ around 22 K – the temperature around which population of the exchange-split excited state of Kramers doublet has been known to occur. This finding suggests that this compound presents a novel situation in which multiferroicity is induced by an interplay between crystal-field effects and exchange interaction. No multiglass features could be observed down 2 K unlike in many other members of this family.

## 1. Introduction

The rare-earth (R) family, $R_2BaNiO_5$, crystallizing in an orthorhombic structure (space group: *Immm*), has been considered to be the first convincing example [1] for Haldane spin-chain behavior and the spin-chain anomalies arise from Ni (S= 1). While these compounds attracted considerable attraction of the community over several years from the magnetism angle [see, for instances, Refs. 2-9], the highly insulating nature of these materials somehow escaped the attention of the community to search for magnetoelectric coupling effects. Therefore, we initiated exhaustive investigations in this direction [10-16], and observed interesting dielectric permittivity and magnetoelectric coupling effects in the materials $R=$ Nd, Gd, Tb, Dy, Ho, and Er. It was observed that some members of this family (R= Gd, Dy, Ho and Er) exhibit ferroelectricity either at Néel temperature ($T_N$) [10, 13, 17] or, due to short-range magnetic correlations, above respective $T_N$ (= ~ 32 − 58 K) [15]. Through band structure calculations [15], it was shown that the centrosymmetric *Immm* structure in the paramagnetic state gets transformed into noncentrosymmetric *Imm2* space group in the magnetically ordered state (due to small local distortions of Ni-O bond distances), thereby favouring ferroelectricity. Glassy features in magnetization (*M*) and complex dielectric permittivity due to magnetoelectric cross-coupling at another lower temperature (<20 K) were also observed. On the other hand, no cross-coupling between magnetic and electric dipoles could be found at $T_N$ (63 K) or above for the Tb member [16]; instead, multiferroic behavior sets in below 25 K which is far below $T_N$ (63 K), without any glassy features in magnetization and dielectric constant ($\varepsilon'$) with an exceptional magnitude of magnetodielectric (MDE) coupling within this series. This extraordinary MDE was attributed to unusual role played by the single-ion anisotropy of trivalent Tb ions. We have also undertaken such studies on magnetoelectric coupling behavior for light rare-earth members. In the case of Nd member ($T_N$= 48 K), no spontaneous electric polarization could be detectable till now above 2 K [14], despite the existence of magnetodielectric coupling. This situation prompted us to investigate other light R members for the cross-coupling effects. In this article, we report the results of such investigations for $Sm_2BaNiO_5$. Our results suggest the absence of ferroelectricity in the vicinity of $T_N$ (55 K), however, revealing bias-pyrocurrent features at a lower temperature (~22 K) as in the case of Tb compound, following exchange splitting of Kramers doublet. This finding brings out an interesting interplay between crystal-field effect, exchange effect and magnetoelectric coupling. No glassy features could be detected in magnetization and dielectric constants data unlike in many other members of this family.

In the family of $R_2BaNiO_5$, the Sm compound is of special interest, as very little is known about the properties of this compound. There is no feature in the temperature dependence of magnetic susceptibility ($\chi$) around $T_N$. However, a sharp peak around 22 K in $\chi(T)$ was reported by Garcia-Matres et al long ago [6], which was subsequently established to arise from Schottky anomaly due to crystal-field effects (see below). Long range magnetic ordering could actually be inferred from the optical spectroscopic studies [9, 18] only from the splitting of the line at 6533 cm$^{-1}$ of the Er$^{3+}$, doped in the compound, and this study revealed the anisotropy of the transferred magnetic field at the Sm site. However, the details associated with the ground state Kramers doublet were reported only recently by Galkin and Klimin [19] through optical spectroscopic studies in the region of the $^6H_{5/2} \rightarrow {}^6F_{3/2}$ intermultiplet transition of trivalent Sm. These authors estimated that the magnetic moment on Sm falls in the range $0.57 − 0.77$ $\mu_B$/Sm and could reproduce the peak in $\chi(T)$ around 22 K taking into account the both Ni and first excited 4f doublet of Sm contributions; it was also shown that the degeneracy of the doublet ground state is lifted by internal magnetic fields below $T_N$ due to a possible complex magnetic ordering from Sm and Ni magnetic fields. Exact nature of the magnetic structure is yet to emerge.

## 2. Experimental details

The polycrystalline form of the compound under investigation was prepared by a standard solid state reaction route. Stoichiometric amounts of $Sm_2O_3$ (purity 99.998%, LEICO Industries), NiO (purity 99.998%, LEICO Industries) and $BaCO_3$ (purity 99.997%, Alfa Aesar) were mixed together thoroughly and heated in air at 900 $^0$C, 1000$^0$C, 1100 $^0$C and 1200 $^0$C for 12 hours each with intermediate grindings and pelletizing. The sample was characterized by x-ray diffraction (Cu K$_\alpha$) and Rietveld fitted pattern is shown in Fig. 1. Dc $\chi(T)$ measurements (1.8-300 K) were measured in magnetic-fields (*H*) of 100 Oe and 5 kOe with the help of a commercial (Quantum Design, USA) magnetometer. The same magnetometer was used to measure isothermal $M$ at selected temperatures. Ac $\chi$ measurements were also performed (1.8 – 120 K) with an ac field of 1 Oe at several frequencies ($\nu$= 1.3, 13, 133 and 1333 Hz). Temperature dependence (1.8 – 120 K) of heat-capacity ($C$) was obtained with a commercial (Quantum Design, USA) Physical Property Measurements System (PPMS) and the same PPMS was used to measure dielectric permittivity down to 1.8 K with the help of a home-made insert; isothermal $H$-dependence of $\varepsilon'$ was measured at some selected temperature. Unless otherwise stated, all the measurements were carried out after cooling the sample from the paramagnetic state (e.g., 150 K) to the desired temperature in zero field. Additionally, we carried out bias electric-field ($E$) measurements with the help of a Keithley 6517B electrometer. For such bias-$E$ measurements, we cooled the sample to 1.8 K in the absence of an electric-field and then the bias current, $I_B$, was measured with the heating rate rate of 2 K/min in the presence of a bias electric field of 2 kV/cm. In addition, pyrocurrent ($I_{pyro}$) behaviour was measured as a function of $T$ while warming in the absence of electric field, after cooling in the presence of an electric field of 2 kV/cm.

### 3. Results and discussion

The results of dc magnetization results are shown in figure 2. The $\chi(T)$, measured in the fields of 100 Oe (inset of Fig. 2a) and 5 kOe (mainframe of figure 2), undergoes a monotonic increase with decreasing $T$, attaining a peak at 22 K and a decrease at lower temperatures. There is another upturn below about 10 K, and it is not clear whether this is due to the tendency for another long range magnetic order below 2 K or from traces of magnetic impurities. No feature due to magnetic ordering could be seen in the plot of $\chi(T)$ around $T_N$. Isothermal magnetization plots at 2, 15 and 40 K (see figure 2b) are found to be linear in the field-range of measurement (<70 kOe) without any evidence for hysteresis. In other words, there is no evidence for any $H$-induced magnetic transitions and this property makes the magnetism of this compound different from many members of this series [10, 12-16]. These features are in good agreement with those reported in the literature [6, 19]. As shown in the inset of figure 2a, there is no bifurcation of the curves obtained for the zero-field-cooled (ZFC) and field-cooled (FC) conditions and the curves follow each other in the features (almost overlapping) down to 1.8 K, without any levelling of FC $\chi$ at low temperatures. This is a sufficient evidence to conclude that this compound does not undergo spin-glass freezing in the measured temperature range. In support of this conclusion, we show the results of ac $\chi$ measurements in Figure 3a. We do not find any cusp in the real part ($\chi'$) that is attributable to spin-glass freezing and the curves obtained with different $\nu$ are found to overlap. In addition, the imaginary part ($\chi''$) is featureless above 1.8 K. We have also measured isothermal remnant magnetization ($M_{IRM}$) at 5 K; for this purpose, we cooled the specimen to 5 K in the absence of an external field, then switched on a field 5 kOe, and after 5 mins, the field was switched off. $M_{IRM}$ was then measured as a function of time. It was found that $M_{IRM}$ falls to negligibly small values within a few seconds after switching off the field and remains constant thereafter, rather than decaying slowly with time (see inset of figure 3a). All these results – absence of a cusp in ZFC-FC dc $\chi$ curves and of bifurcation of these curves, absence of hysteresis in isothermal $M$, absence of frequency dispersion in ac $\chi$, featureless $\chi''$ and of decay of $M_{IRM}$ - conclusively rule out spin-glass freezing.

As stated earlier, barring optical studies, there was no experimental evidence for the onset of long range magnetic order. We have therefore measured heat-capacity as a function of temperature. The plot of $C(T)$ exhibits an upturn below 52 K, attributable to the onset of long range magnetic order, exhibiting a weak peak near 48 K (Fig. 3b). This is more evident in the plot of $C/T$. A shoulder in $C(T)$ also appears near 22 K due to Schottky anomaly, which is distinctly seen as a peak in the plot of $C/T$. A weak upturn in $C/T$ (though not transparent in the plot of $C$ versus $T$) below about 5 K may be compared with a similar feature in $\chi(T)$ and future investigations may have to address whether it is intrinsic.

In figure 4a, we show the real part ($\varepsilon'$) of complex dielectric permittivity as a function of $T$, measured with various frequencies and 1V ac bias. Since there is no worthwhile feature above 40 K, we show the data below this temperature only. We noted that $\varepsilon'$ decreases monotonically as the temperature is decreased from 300 K, with the values being 4012 and 15 at 300 and 60 K respectively; however, below about 60 K, there is a weak increase without any well-defined feature at $T_N$, resulting in a weak peak at 22 K followed by a decrease. The absence of a well-defined feature at $T_N$ implies absence of a coupling between magnetic and electric dipoles at the onset of long-range magnetic order.

Interestingly, this cross-coupling is triggered after the splitting of the Kramers doublet ground state by the exchange field (around 22 K). [We would like to mention that dielectric anomalies due to crystal-field effects was seen even in the Nd case, Ref. 14]. In order to confirm this, we have also measured magnetodielectric effect by measuring the change in $\varepsilon'$ with the application of external magnetic field. The plot of $\Delta\varepsilon'(H)$ (= {$\varepsilon'(H)$-$\varepsilon'(0)$}/$\varepsilon'(0)$) is shown in figure 4b at selected temperatures. As seen in this figure, above 22 K, e.g., at 30 K, $\Delta\varepsilon'$ is practically zero, whereas the values are sufficiently significant at high fields at 2 and 15 K. Clearly, MDE is triggered well below 22 K only.     It is to be noted that there is no frequency dispersion of the peak temperature  and in fact the $\varepsilon'(T)$ curves actually overlap after normalizing to respective peak values, as shown in the inset of figure 4a. Thus, no glassy feature in $\varepsilon'(T)$ can be traced in our data. It may be added that the values of the loss part (tan$\delta$) fall in the range $0.05 - 2$ above 60 K, but  are very small below 60 K (0.0003, 0.001 and 0.007 at 2, 20 and 50 K respectively). Therefore, the compound is sufficiently insulating that extrinsic contributions like leakage current can be ruled out in the $T$-range of interest.

In figure 5a, we show $I_{pyro}$  as a function of $T$.  We see a distinct peak in the vicinity of $T_N$. The sign of $I_{pyro}$ could be reversed  by changing the polarity of the electric field, as shown in the inset of figure 5a. These appear to arise from the onset of ferroelectricity around $T_N$. However, if the $I_{pyro}$ is recorded for different heating rates, the peak temperature shifts, e.g., from 50 K for 2 K/min to 54 K for 4 K/min to 56 K for 6K/min. Such a rate dependence is a highly persuasive evidence against any interpretation in terms of ferroelectricity in the vicinity of $T_N$ and is usually interpreted in terms of 'thermally stimulated depolarization current' [20-26]. In order to gain further support to this conclusion, we followed a procedure suggested in the recent literature [22, 27], that is, to measure bias current as mentioned in the experimental section.  We found that the $I_B$ peak  in the vicinity of $T_N$ is not seen and $I_B$ increases smoothly across $T_N$, even without showing a shoulder (see the inset of figure 5b). However, such a pyrocurrent curve obtained in the presence of an electric field reveals a weak peak around 22 K, reversing its sign for the opposite polarity of the electric field, as shown in figure 5b. This is the same temperature at which a peak in $\varepsilon'(T)$ as well as in $\chi(T)$ appears.  This is an intriguing finding bringing out the requirement of splitting of Kramers doublet for the observation of  multiferroicity in the magnetically ordered state.

## 4.  Summary

We have reported magnetic, dielectric and pyrocurrent behavior of the Haldane spin-chain compound, $Sm_2BaNiO_5$. In contrast to what was reported for Gd, Dy, Ho and Er members, this compound does not exhibit magnetoelectric coupling due to the onset of long range magnetic order. Instead, as indicated by bias pyroelectric data, cross-coupling is induced following the splitting of the Kramers doublet. Thus this work brings an intriguing role of single-ion anisotropy following an interplay between crystal-field splitting, and exchange-effect on magnetoelectric coupling. This compound is in addition different from above R members in the sense that  multiglass features are not seen  in the former.

Finally, we would like to state that the pyrocurrent measured in the presence of a bias electric-field on $Nd_2BaNiO_5$ reveals a peak  around 10 K (see figure 6), confirming the existence of ferroelectric features following the lifting of degeneracy of the (crystal-field-split) doublet ground state by the exchange field in this compound as well. This supports the idea proposed on the basis of the studies on the Sm compound.

Figure captions:

Figure 1:
X-ray diffraction of $Sm_2BaNiO_5$ at room temperature, along with Reitveld refinement.

Figure 2:
For $Sm_2BaNiO_5$, (a) temperature dependence of magnetic susceptibility measured in a field of 5 kOe and (b) isothermal magnetization behavior at 2, 15 and 40 K. Magnetic susceptibility, obtained in a field of 100 Oe for the zero-field-cooled and field-cooled conditions of the specimen, are plotted below 100 K in the inset of (a). The lines through the data points serve as guides to the eyes.

Figure 3:
For $Sm_2BaNiO_5$, (a) real and imaginary parts of ac susceptibility as a function of temperature (2 – 60 K) and (b) heat-capacity and heat-capacity divided by temperature below 80 K. Isothermal remnant magnetization behavior at 5 K is plotted as a function of time ($t$) in the inset of (a). In figure (a), lines are drawn through the data points.

Figure 4:
For $Sm_2BaNiO_5$, (a) dielectric constant as a function of temperature below 40 K at various frequencies, and (b) the change in the dielectric constant with magnetic field at 2, 15 and 30. In the inset of (a), the curves are normalized to respective peak values of $\varepsilon'$ to show that these curves overlap and are dispersionless. Lines are drawn through the data points.

Figure 5:
For $Sm_2BaNiO_5$, (a) pyroelectric current as a function of temperature (<90 K) for different rates of heating – 2, 4, 6 K/min , and (b) bias pyroelectric current for two opposite polarities of electric field obtained as described in the text. In the inset of (a) the reversal of the sign of pyrocurrent for the change of polarity of electric field is shown. In the inset of (b), bias pyrocurrent behavior in the vicinity of $T_N$ (50 – 60 K) is shown to bring out that there is no feature in this $T$-range. Lines are drawn through the data points.

Figure 6:
Bias pyrocurrent behavior for two opposite polarities of electric field in $Nd_2BaNiO_5$.

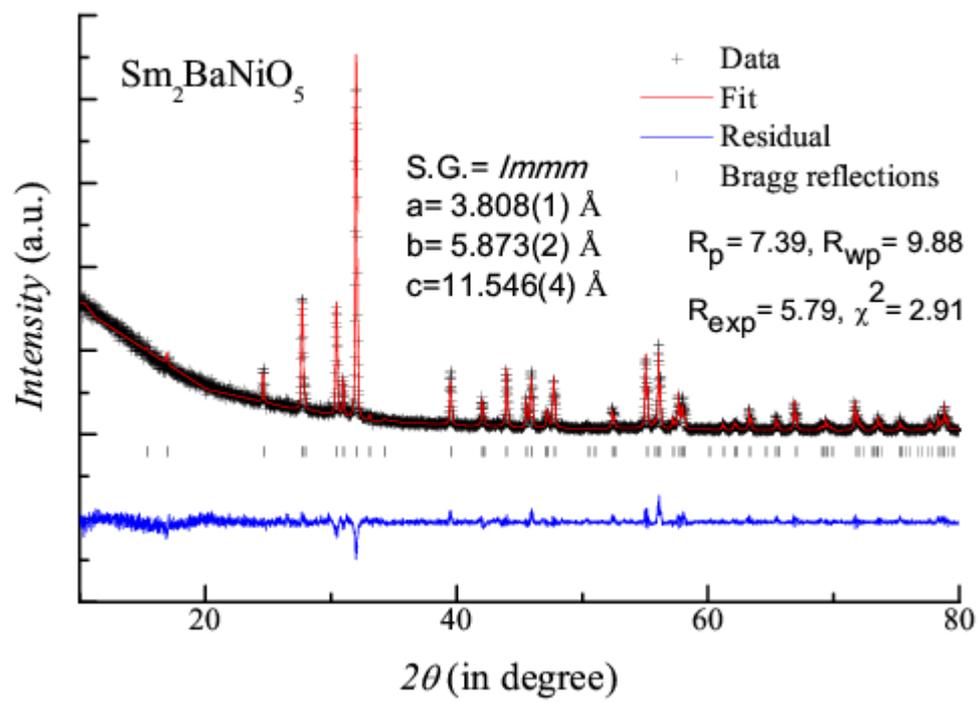

Figure 1:

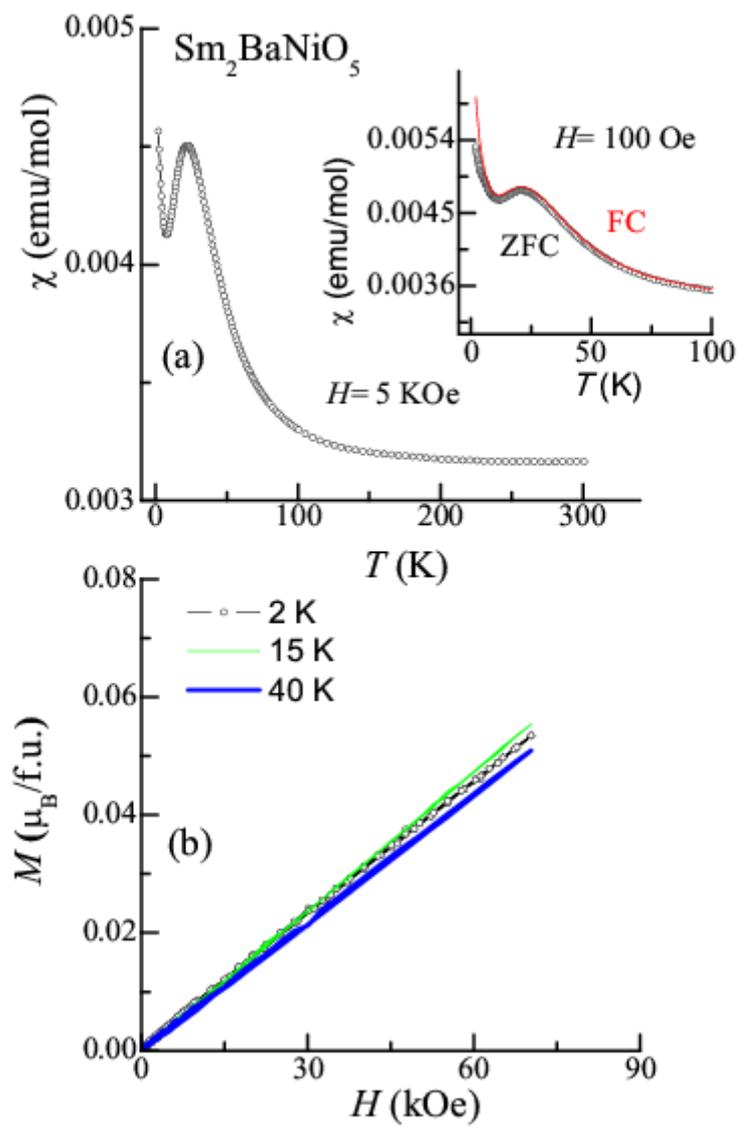

Figure 2:

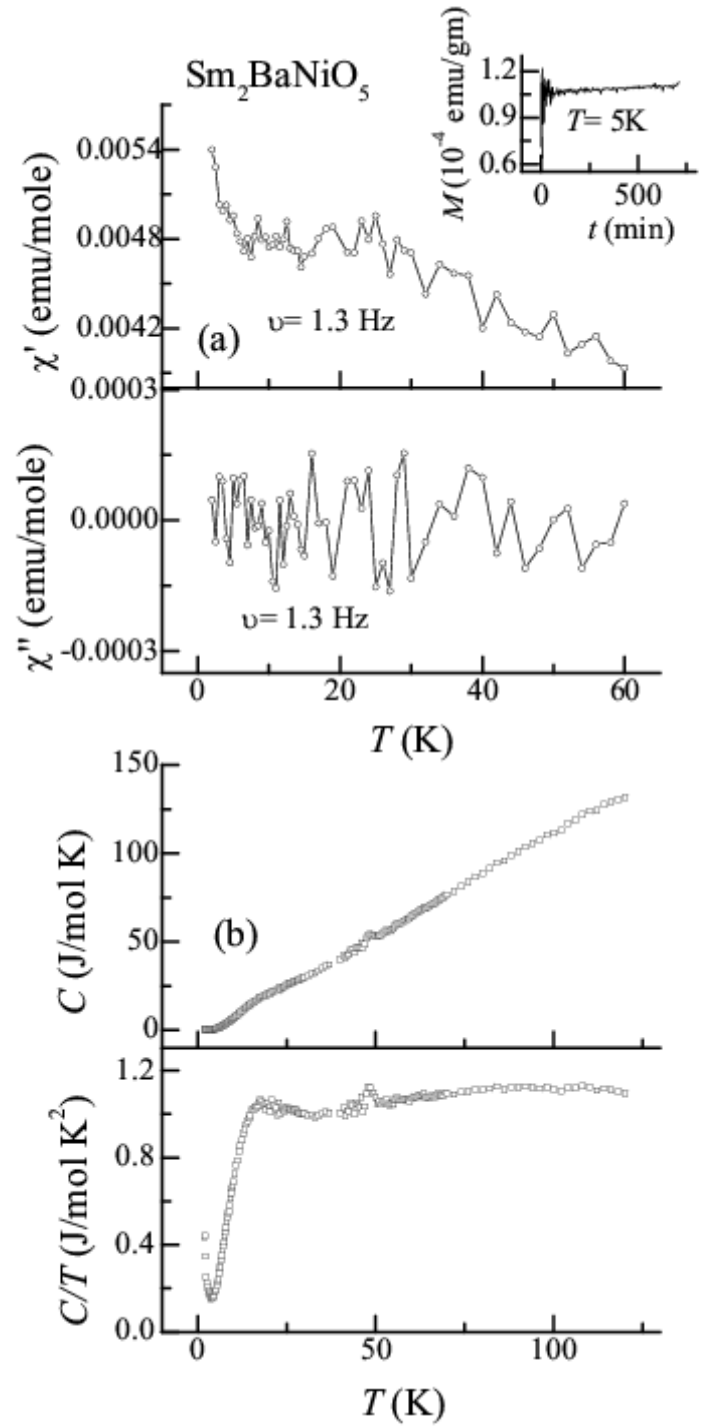

Figure 3:

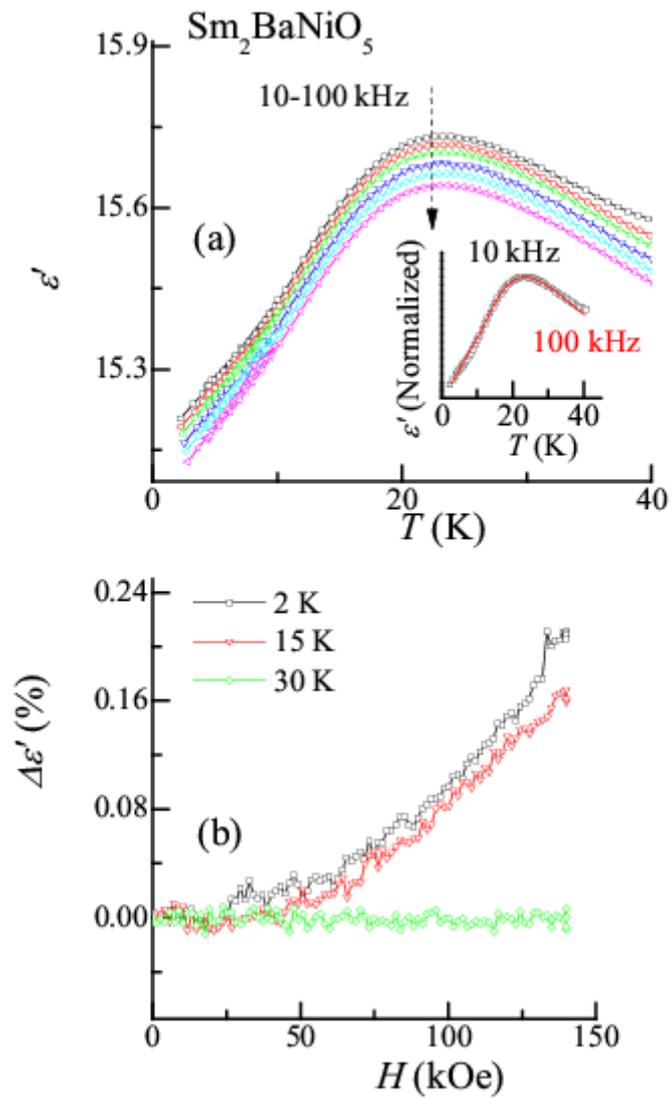

Figure 4:

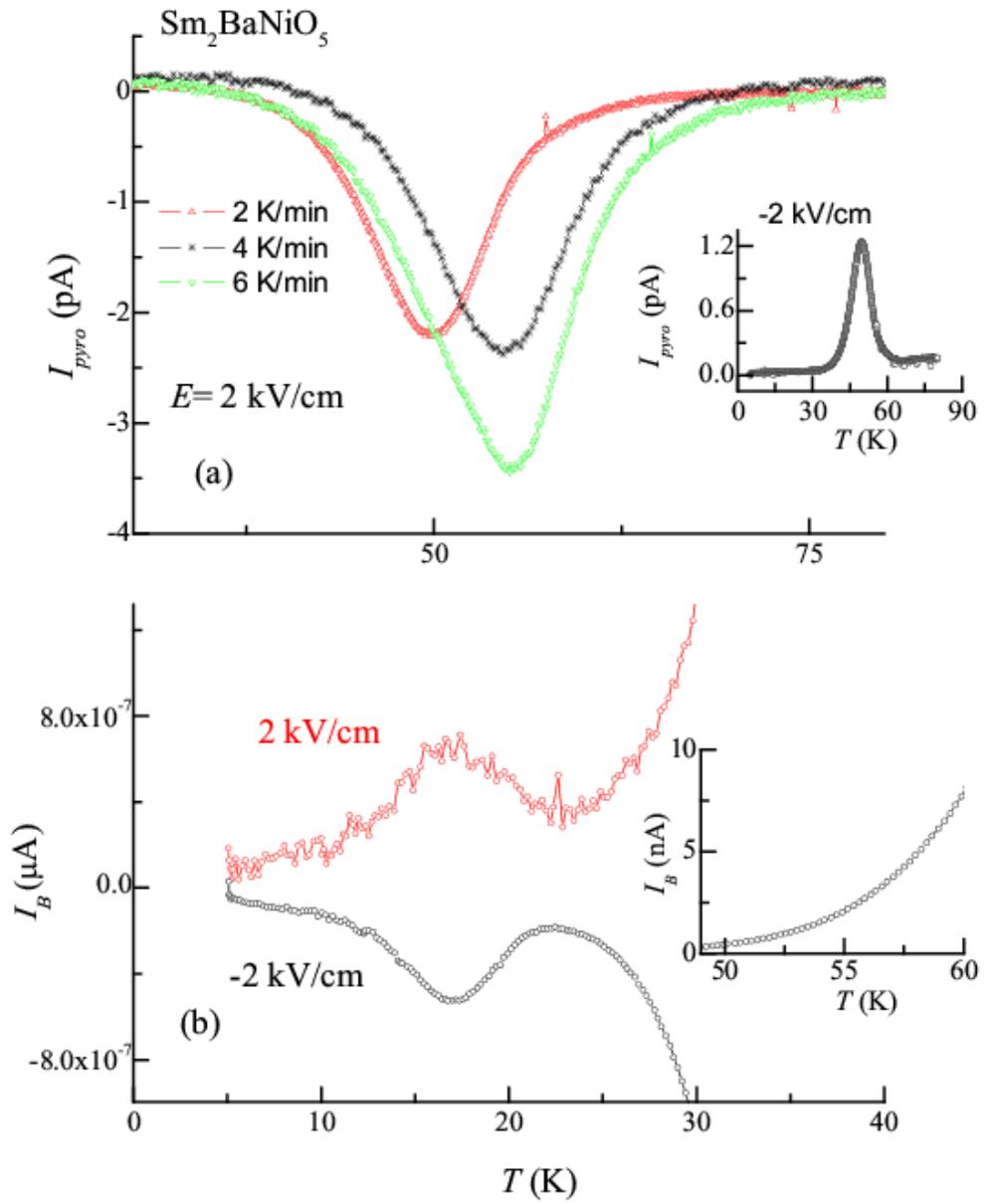

Figure 5:

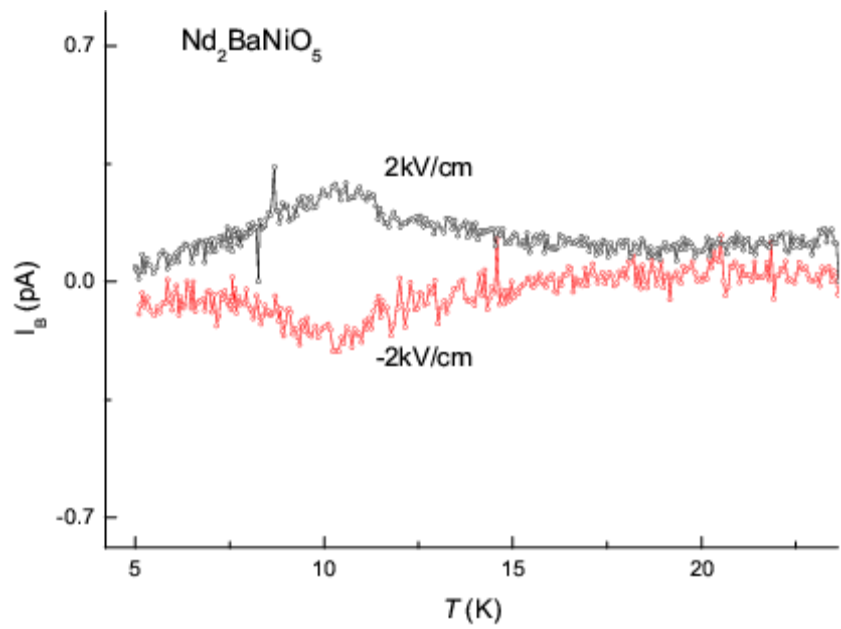

Figure 6